\RequirePackage{lineno}
\documentclass[twocolumn,twoside]{revtex4}
\usepackage{graphicx}
\usepackage{fancyhdr}
\usepackage{upgreek} 
\usepackage{ifthen} 
\newboolean{uprightparticles}
\setboolean{uprightparticles}{false} 

\usepackage{xspace} 
\usepackage{upgreek}

\def\lhcb {\mbox{LHCb}\xspace}

\def\babar  {\mbox{BaBar}\xspace}
\def\belle  {\mbox{Belle}\xspace}

\def\MagUp {\mbox{\em Mag\kern -0.05em Up}\xspace}

\ifthenelse{\boolean{uprightparticles}}%
{

 \def\Pmu         {\ensuremath{\upmu}\xspace}                 
 \def\Pnu         {\ensuremath{\upnu}\xspace}                 
                  
 \def\Ppi         {\ensuremath{\uppi}\xspace}

 \def\Ptau        {\ensuremath{\uptau}\xspace}

 \def\Ppsi        {\ensuremath{\uppsi}\xspace}

 \def\PDelta      {\ensuremath{\Delta}\xspace}                 
 \def\PXi      {\ensuremath{\Xi}\xspace}                 
 \def\PLambda      {\ensuremath{\Lambda}\xspace}                 
 \def\PSigma      {\ensuremath{\Sigma}\xspace}                 
 \def\POmega      {\ensuremath{\Omega}\xspace}                 
 \def\PUpsilon      {\ensuremath{\Upsilon}\xspace}                 
 
 \def\PB      {\ensuremath{\mathrm{B}}\xspace}                 
                  
 \def\PD      {\ensuremath{\mathrm{D}}\xspace}

 \def\PJ      {\ensuremath{\mathrm{J}}\xspace}                 
 \def\PK      {\ensuremath{\mathrm{K}}\xspace}

 \def\Pb      {\ensuremath{\mathrm{b}}\xspace}                 
 \def\Pc      {\ensuremath{\mathrm{c}}\xspace}                 
                  
 \def\Pe      {\ensuremath{\mathrm{e}}\xspace}

 \def\Pi      {\ensuremath{\mathrm{i}}\xspace}

 \def\Pp      {\ensuremath{\mathrm{p}}\xspace}

 \def\Ps      {\ensuremath{\mathrm{s}}\xspace}

}
{

 \def\Pmu         {\ensuremath{\mu}\xspace}                 
 \def\Pnu         {\ensuremath{\nu}\xspace}                 
                  
 \def\Ppi         {\ensuremath{\pi}\xspace}

 \def\Ptau        {\ensuremath{\tau}\xspace}

 \def\Ppsi        {\ensuremath{\psi}\xspace}                 
                  
 \mathchardef\PDelta="7101
 \mathchardef\PXi="7104
 \mathchardef\PLambda="7103
 \mathchardef\PSigma="7106
 \mathchardef\POmega="710A
 \mathchardef\PUpsilon="7107
                  
 \def\PB      {\ensuremath{B}\xspace}                 
                  
 \def\PD      {\ensuremath{D}\xspace}

 \def\PJ      {\ensuremath{J}\xspace}                 
 \def\PK      {\ensuremath{K}\xspace}

 \def\Pb      {\ensuremath{b}\xspace}                 
 \def\Pc      {\ensuremath{c}\xspace}                 
                  
 \def\Pe      {\ensuremath{e}\xspace}

 \def\Pi      {\ensuremath{i}\xspace}

 \def\Pp      {\ensuremath{p}\xspace}

 \def\Ps      {\ensuremath{s}\xspace}

}

\DeclareRobustCommand{\optbar}[1]{\shortstack{{\miniscule (\rule[.5ex]{1.25em}{.18mm})}
  \\ [-.7ex] $#1$}}

\def\epem       {{\ensuremath{\Pe^+\Pe^-}}\xspace}

\def\mup        {{\ensuremath{\Pmu^+}}\xspace}
\def\mun        {{\ensuremath{\Pmu^-}}\xspace} 

\def\taup       {{\ensuremath{\Ptau^+}}\xspace}
\def\taum       {{\ensuremath{\Ptau^-}}\xspace}

\def\ellm       {{\ensuremath{\ell^-}}\xspace}
\def\ellp       {{\ensuremath{\ell^+}}\xspace}
\def\ellell     {\ensuremath{\ell^+ \ell^-}\xspace}

\def\neu        {{\ensuremath{\Pnu}}\xspace}
\def\neub       {{\ensuremath{\overline{\Pnu}}}\xspace}
\def\neum       {{\ensuremath{\neu_\mu}}\xspace}
\def\neumb      {{\ensuremath{\neub_\mu}}\xspace}
\def\neut       {{\ensuremath{\neu_\tau}}\xspace}
\def\neutb      {{\ensuremath{\neub_\tau}}\xspace}
\def\neul       {{\ensuremath{\neu_\ell}}\xspace}
\def\neulb      {{\ensuremath{\neub_\ell}}\xspace}

\def\squark    {{\ensuremath{\Ps}}\xspace}

\def\cquark    {{\ensuremath{\Pc}}\xspace}

\def\bquark    {{\ensuremath{\Pb}}\xspace}

\def\pion   {{\ensuremath{\Ppi}}\xspace}
\def\piz    {{\ensuremath{\pion^0}}\xspace}

\def\pip    {{\ensuremath{\pion^+}}\xspace}
\def\pim    {{\ensuremath{\pion^-}}\xspace}

  \def\Kbar    {{\kern 0.2em\overline{\kern -0.2em \PK}{}}\xspace}

\def\KorKbar    {\kern 0.18em\optbar{\kern -0.18em K}{}\xspace}

  \def\Dbar    {{\kern 0.2em\overline{\kern -0.2em \PD}{}}\xspace}
\def\D       {{\ensuremath{\PD}}\xspace}

\def\DorDbar    {\kern 0.18em\optbar{\kern -0.18em D}{}\xspace}
\def\Dz      {{\ensuremath{\D^0}}\xspace}

\def\Dp      {{\ensuremath{\D^+}}\xspace}

\def\Dstar   {{\ensuremath{\D^*}}\xspace}

\def\Dstarz  {{\ensuremath{\D^{*0}}}\xspace}

\def\Dstarp  {{\ensuremath{\D^{*+}}}\xspace}
\def\Dstarm  {{\ensuremath{\D^{*-}}}\xspace}

\def\Ds      {{\ensuremath{\D^+_\squark}}\xspace}
\def\Dsp     {{\ensuremath{\D^+_\squark}}\xspace}

\def\B       {{\ensuremath{\PB}}\xspace}
\def\Bbar    {{\ensuremath{\kern 0.18em\overline{\kern -0.18em \PB}{}}}\xspace}
\def\Bb      {{\ensuremath{\Bbar}}\xspace}
\def\BorBbar    {\kern 0.18em\optbar{\kern -0.18em B}{}\xspace}
\def\Bz      {{\ensuremath{\B^0}}\xspace}
\def\Bzb     {{\ensuremath{\Bbar{}^0}}\xspace}
\def\Bu      {{\ensuremath{\B^+}}\xspace}
\def\Bub     {{\ensuremath{\B^-}}\xspace}
\def\Bp      {{\ensuremath{\Bu}}\xspace}
\def\Bm      {{\ensuremath{\Bub}}\xspace}

\def\Bs      {{\ensuremath{\B^0_\squark}}\xspace}

\def\Bdb     {{\ensuremath{\Bbar{}^0}}\xspace}
\def\Bc      {{\ensuremath{\B_\cquark^+}}\xspace}

\def\jpsi     {{\ensuremath{{\PJ\mskip -3mu/\mskip -2mu\Ppsi\mskip 2mu}}}\xspace}

  \def\Y#1S{\ensuremath{\PUpsilon{(#1S)}}\xspace}

\def\FourS {{\Y4S}\xspace}

\def\proton      {{\ensuremath{\Pp}}\xspace}

\def\Lz          {{\ensuremath{\PLambda}}\xspace}
\def\Lbar        {{\ensuremath{\kern 0.1em\overline{\kern -0.1em\PLambda}}}\xspace}
\def\LorLbar    {\kern 0.18em\optbar{\kern -0.18em \PLambda}{}\xspace}

\def\Lb      {{\ensuremath{\Lz^0_\bquark}}\xspace}

\def\Lc      {{\ensuremath{\Lz^+_\cquark}}\xspace}

\def\BF         {{\ensuremath{\mathcal{B}}}\xspace}

\def\BR         {\BF}

\def\to                 {\ensuremath{\rightarrow}\xspace}

\def\qsq       {{\ensuremath{q^2}}\xspace}

\def\AT#1     {\ensuremath{A_{\mathrm{T}}^{#1}}\xspace}           

\def\C#1      {\ensuremath{\mathcal{C}_{#1}}\xspace}                       
\def\Cp#1     {\ensuremath{\mathcal{C}_{#1}^{'}}\xspace}                    
\def\Ceff#1   {\ensuremath{\mathcal{C}_{#1}^{\mathrm{(eff)}}}\xspace}        
\def\Cpeff#1  {\ensuremath{\mathcal{C}_{#1}^{'\mathrm{(eff)}}}\xspace}       
\def\Ope#1    {\ensuremath{\mathcal{O}_{#1}}\xspace}                       
\def\Opep#1   {\ensuremath{\mathcal{O}_{#1}^{'}}\xspace}                    

\newcommand{\tev}{\ifthenelse{\boolean{inbibliography}}{\ensuremath{~T\kern -0.05em eV}}{\ensuremath{\mathrm{\,Te\kern -0.1em V}}}\xspace}
\newcommand{\gev}{\ensuremath{\mathrm{\,Ge\kern -0.1em V}}\xspace}
\newcommand{\mev}{\ensuremath{\mathrm{\,Me\kern -0.1em V}}\xspace}
\newcommand{\kev}{\ensuremath{\mathrm{\,ke\kern -0.1em V}}\xspace}
\newcommand{\ev}{\ensuremath{\mathrm{\,e\kern -0.1em V}}\xspace}
\newcommand{\gevc}{\ensuremath{{\mathrm{\,Ge\kern -0.1em V\!/}c}}\xspace}
\newcommand{\mevc}{\ensuremath{{\mathrm{\,Me\kern -0.1em V\!/}c}}\xspace}
\newcommand{\gevcc}{\ensuremath{{\mathrm{\,Ge\kern -0.1em V\!/}c^2}}\xspace}
\newcommand{\gevgevcccc}{\ensuremath{{\mathrm{\,Ge\kern -0.1em V^2\!/}c^4}}\xspace}
\newcommand{\mevcc}{\ensuremath{{\mathrm{\,Me\kern -0.1em V\!/}c^2}}\xspace}

\def\invfb   {\ensuremath{\mbox{\,fb}^{-1}}\xspace}

\newcommand{\stat}{\ensuremath{\mathrm{\,(stat)}}\xspace}
\newcommand{\syst}{\ensuremath{\mathrm{\,(syst)}}\xspace}

\def\gsim{{~\raise.15em\hbox{$>$}\kern-.85em
          \lower.35em\hbox{$\sim$}~}\xspace}
\def\lsim{{~\raise.15em\hbox{$<$}\kern-.85em
          \lower.35em\hbox{$\sim$}~}\xspace}

\def\sqs   {\ensuremath{\protect\sqrt{s}}\xspace}

\def\tell1  {TELL1\xspace}
\def\ukl1   {UKL1\xspace}

\def\RD       {{\ensuremath{\mathcal{R}(\D)}}\xspace}
\def\Hc       {{\ensuremath{H_c}}\xspace}
\def\Hb       {{\ensuremath{H_b}}\xspace}
\def\RHc       {{\ensuremath{\mathcal{R}(\Hc)}}\xspace}
\def\RDst     {{\ensuremath{\mathcal{R}(\Dstar)}}\xspace}

\def\RJpsi     {{\ensuremath{\mathcal{R}(\jpsi)}}\xspace}

\def\mmiss     {{\ensuremath{m_{\rm miss}^2}}\xspace}

\def\qsq      {{\ensuremath{q^2}}\xspace}

\begin{document}
\title{Lepton flavour universality in charged-current \B decays}

\author{S. Klaver\\
on behalf of the LHCb collaboration\\
also including results from the Belle collaboration}
\affiliation{INFN Laboratori Nazionali di Frascati, Via Enrico Fermi, 40, 00044 Frascati, Italy}

\begin{abstract}
Tests of lepton flavour universality in charged-current \B decays offer an excellent opportunity to test the Standard Model,
and show hints of new physics in analyses performed by the \lhcb, \belle and \babar experiments. These proceedings present the
results from the \lhcb collaboration on measurements of \RDst and \RJpsi. It also presents the latest semileptonic tag 
measurement of \RD and \RDst by the \belle collaboration. The latest HFLAV average shows a discrepancy of 3.1$\sigma$
between the Standard Model predictions and combined measurements of \RD and \RDst.
\end{abstract}

\maketitle

\thispagestyle{fancy}

\section{Introduction}
In the Standard Model of particle physics (SM) it is assumed that there are three generations of fermions which 
are nearly identical copies of one another with the same gauge charge assignments, but different masses. 
This implies that all leptons couple universally to the gauge bosons, and that the only difference in their interactions is 
caused by the difference in mass. This is called lepton flavour universality (LFU) and can be tested by measuring ratios
of decays, such that the Cabibbo-Kobayashi-Maskawa matrix elements, and the majority of the form factors, cancel in the ratio. 

These proceedings focus on the measurements of LFU in charged-current
\B~decays, which are of the form $\bquark\to\cquark\ellm\neulb$, commonly known as measurements of \RHc. 
The ratio \RHc is defined as
\begin{equation}
    \RHc = \frac{\BR(\Hb\to\Hc\taum\neutb)}{\BR(\Hb\to\Hc\ell^-\neulb)} \, ,
\end{equation}
where \Hb and \Hc are a \bquark and \cquark hadron, respectively, and $\ell$ is either an electron or muon. 
The semitauonic decay is called the signal channel, and the other decay is the normalisation channel. 
These tree-level processes are theoretically clean and are sensitive to new physics, such as charged
Higgs bosons or leptoquarks~\cite{Buttazzo:2017ixm}.
Up until the start of 2019, there was a discrepancy of 4$\sigma$ between the SM predictions
and the combined measurements of \RD and \RDst.

There are two types of experiments that have measured the ratios \RHc. The first are the \B factories \babar and \belle, which were 
both located at \epem colliders running at the \FourS resonance to produce \Bp\Bm or \Bz\Bzb pairs. They have the advantage 
that \B mesons are produced in a clean environment with little background and that the well-constrained kinematics are very beneficial
for reconstructing final states with neutrinos. The \babar and \belle experiments finished data taking in 2008 and 2010 and collected
433\invfb and 711\invfb of data, respectively. 

LFU in charged-current \B decays can also be measured at the \lhcb experiment, which records data from $pp$ collisions at the LHC. 
The \bquark quarks are produced through gluon fusion and thus all \bquark-hadron species are created: \Bp, \Bz, 
\Bs, \Bc and \Lb. The \bquark hadrons are strongly boosted, providing an excellent separation between production and decay vertices.
However, the large amount of \bquark quarks created comes at the cost of large amounts of background. The \lhcb experiment recorded 3\invfb 
of data in 2011--2012 at \sqs=7--8 TeV (Run 1), and 6\invfb from 2015--2018 at \sqs=13 TeV (Run 2).

\section{Measurements from LHCb}

This section presents \lhcb's three measurements of LFU in charged-current \B decays.

\subsection{Muonic \RDst}

\begin{figure*}
  \centering
  \setbox1=\hbox{\includegraphics[width=\textwidth]{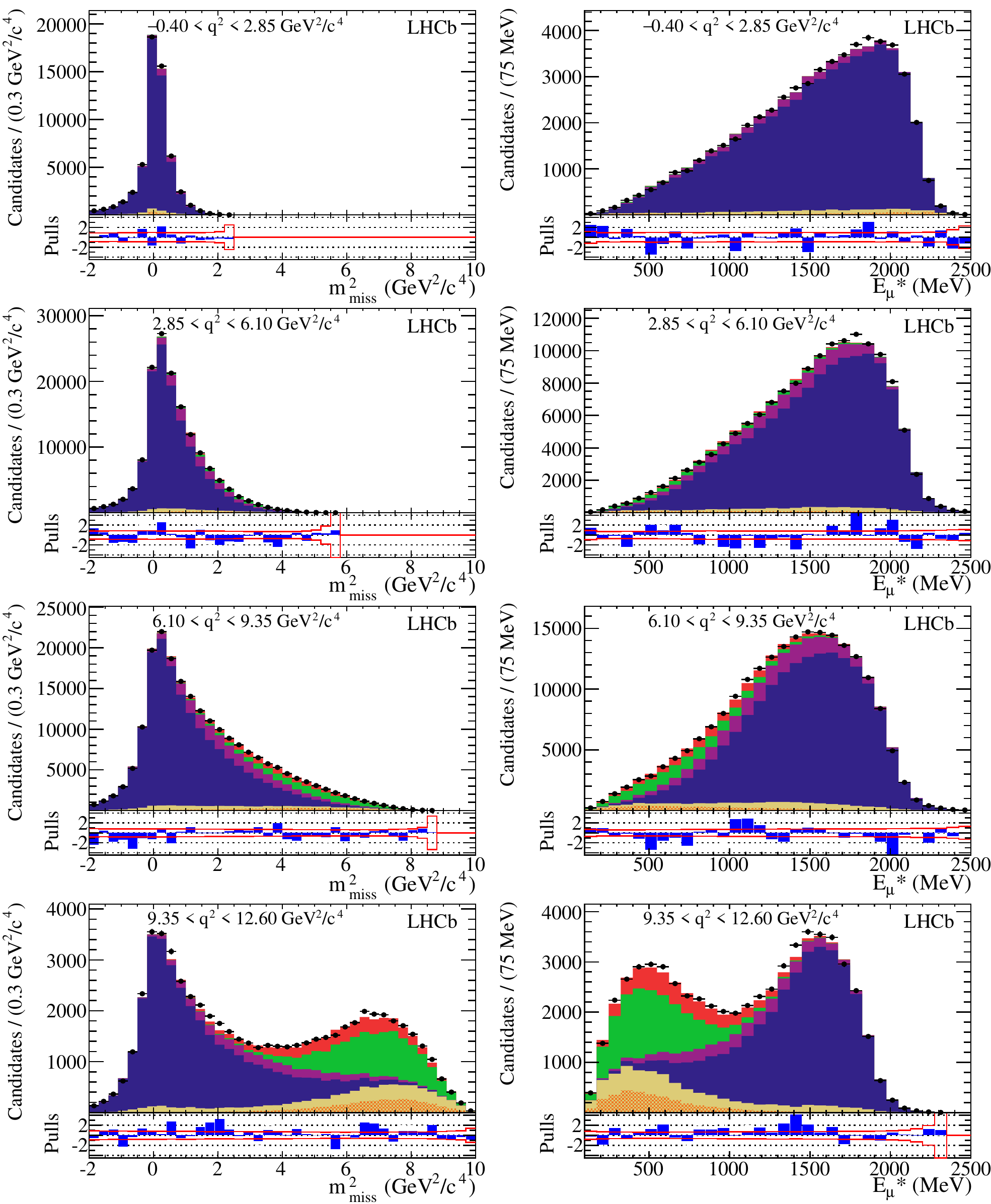}}  
  \includegraphics[width=\textwidth, trim={0 0 0 18.2cm},clip]{figs/figure1.pdf}\llap{\makebox[13cm][l]{\raisebox{3.2cm}{\includegraphics[height=1.45cm]{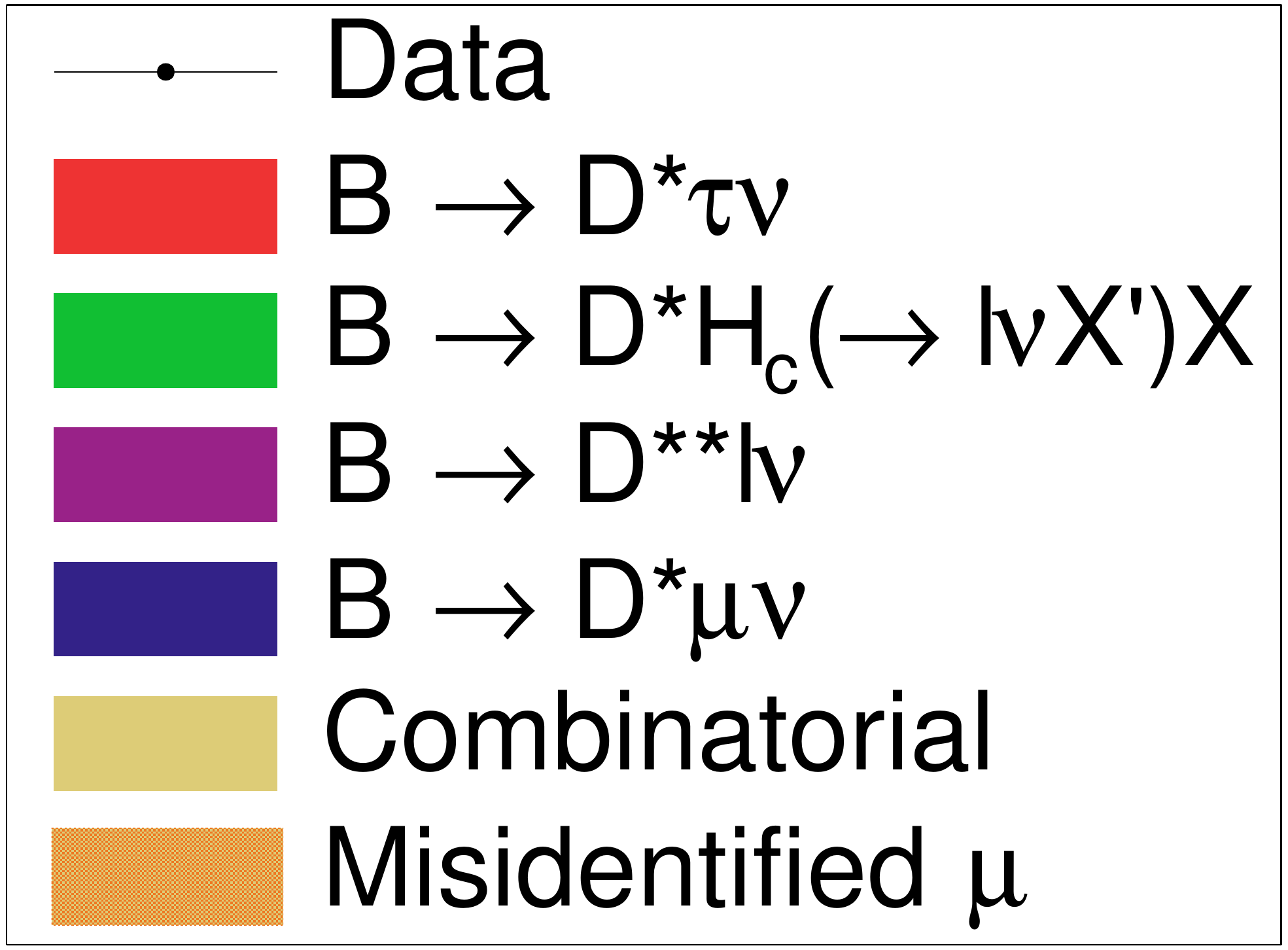}}}}
  \vspace{-0.4cm}
  \caption{\label{fig:RDst_fits} Distributions of $\rm{m}_{\rm{miss}}^2$ (left) and  $E_{\mu}^{*}$ (right) in the highest $q^2$ bin of the signal data, overlaid with the projections of the fit model from \lhcb's muonic \RDst measurement~\cite{Aaij:2015yra}. The signal distributions are red, and the normalisation channel is blue.}
\end{figure*}

The \RDst analysis~\cite{Aaij:2015yra} measure the ratio
\begin{equation}
\RDst = \frac{ \BR( \Bdb \to \Dstarp \taum \neutb ) } { \mathcal{B}( \Bdb \to \Dstarp \mun \neumb ) } \, .
\label{eq:RDstar}
\end{equation}
In this analysis, the \taum decay is reconstructed as \taum \to \mun \neumb \neut, which means that the signal and 
normalisation channel both have the same visible final state. This ensures the cancellation of many systematic uncertainties in 
the ratio, but also makes it hard to distinguish between the two channels. The decay modes are measured using a multidimensional
template fit based on the three kinematic variables that discriminate most between signal and normalisation channels. These are
the missing mass squared (\mmiss), the muon energy ($E_{\mu}^*$) and the squared four-momentum of the lepton pair (\qsq),
all computed in the \B-meson rest frame. An approximation of the boost of the \B meson is made by assuming that the boost of the 
visible decay products along the $z$-axis is equal to that of the \B meson: $(\gamma\beta_z)_{\B} \approx (\gamma\beta_z)_{\Dstar\mu}$

The analysis is performed using the Run 1 data set of \lhcb. The results of the fit, in the highest \qsq bin, are shown in 
Fig~\ref{fig:RDst_fits}. After correcting for the efficiencies of reconstructing the signal and normalisation mode, they yield a value of 
\begin{eqnarray*}
\RDst = 0.336 \pm 0.027\stat \pm 0.030\syst \, .
\end{eqnarray*}
The largest contribution to the systematic uncertainty is due to the limited size of the simulation samples used to create the template
shapes. The obtained value of \RDst is compatible with the SM within 2.1$\sigma$.

\subsection{Hadronic \RDst}

\begin{figure*}[tbp]
\includegraphics[width=0.32\textwidth]{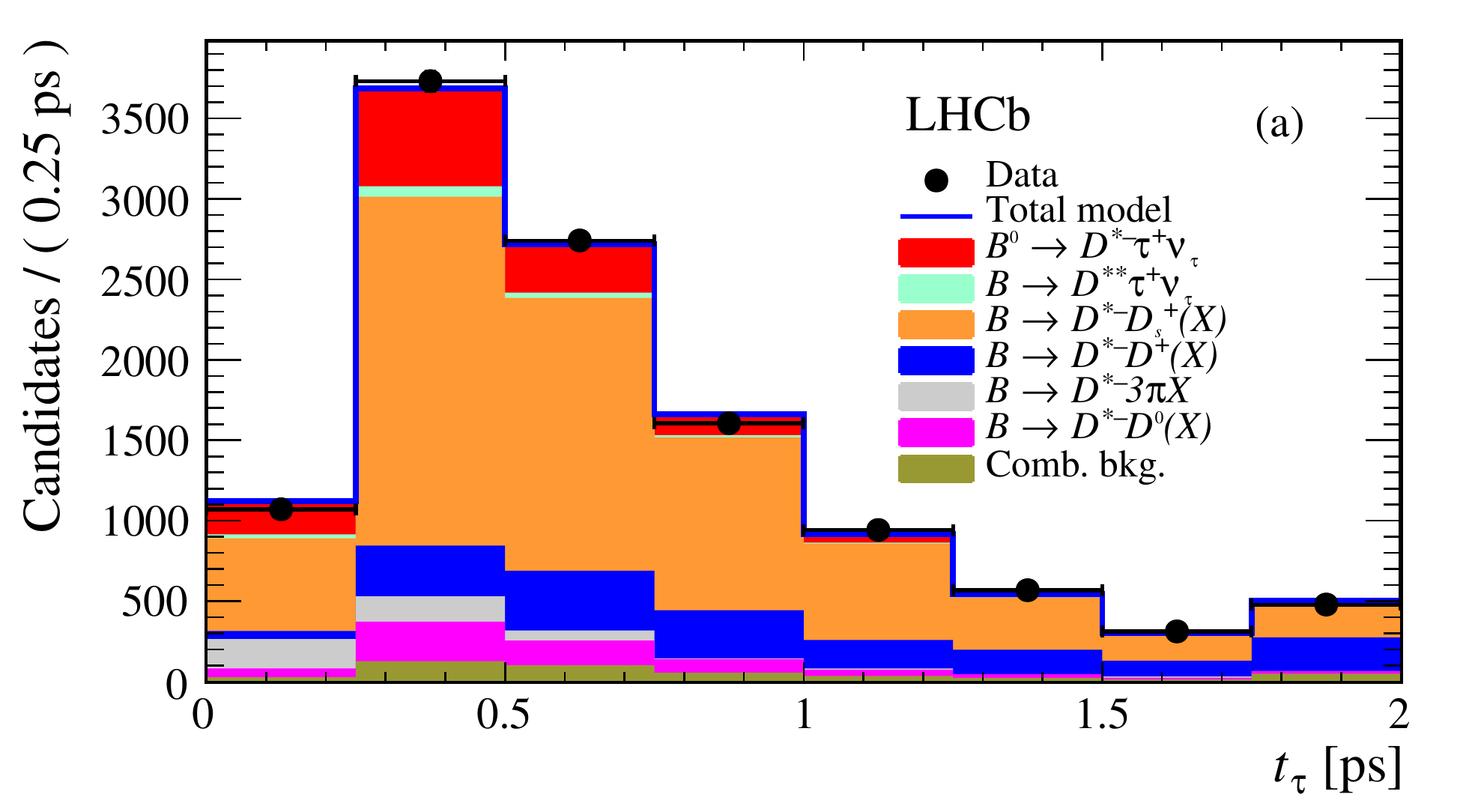}
\includegraphics[width=0.32\textwidth]{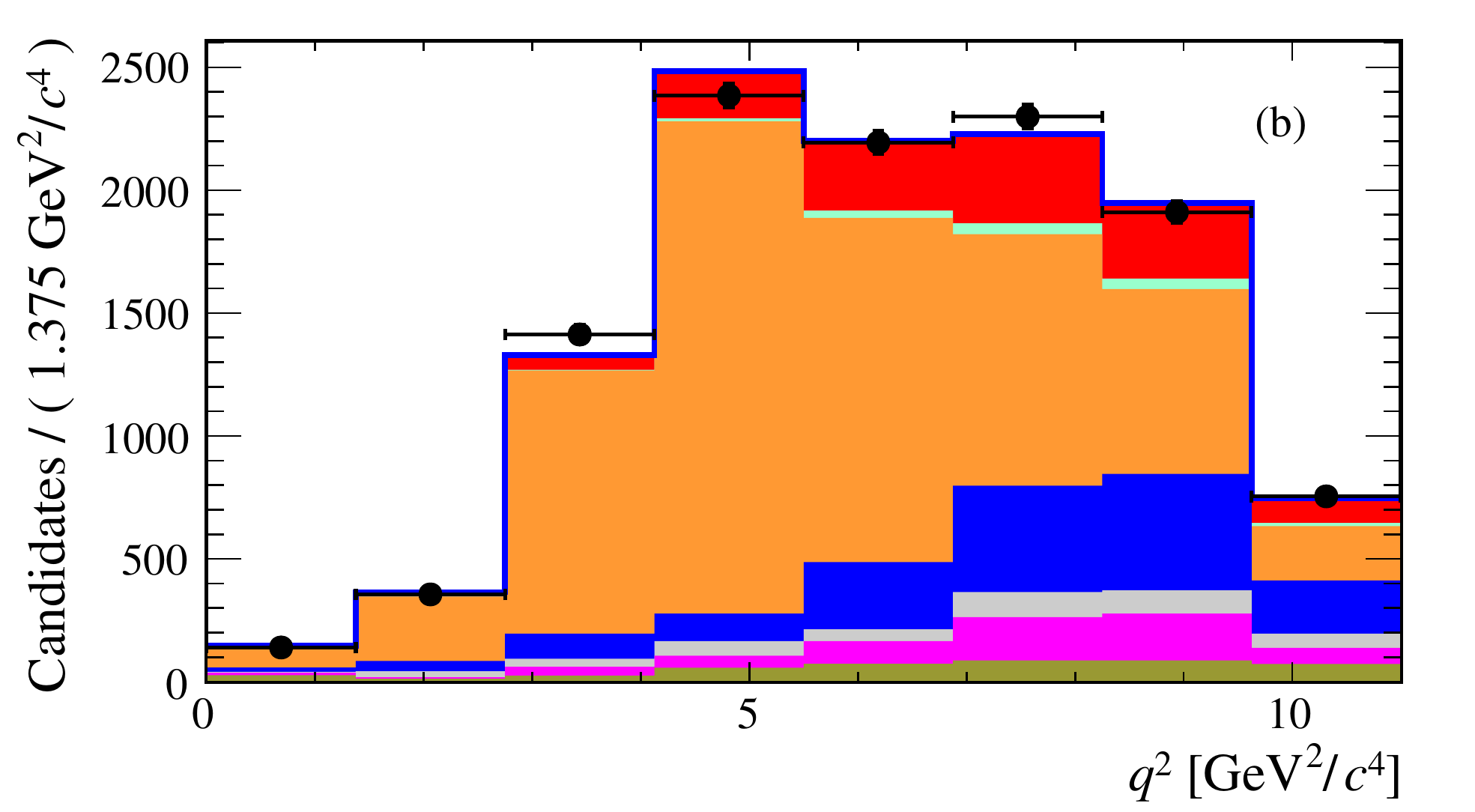}
\includegraphics[width=0.32\textwidth]{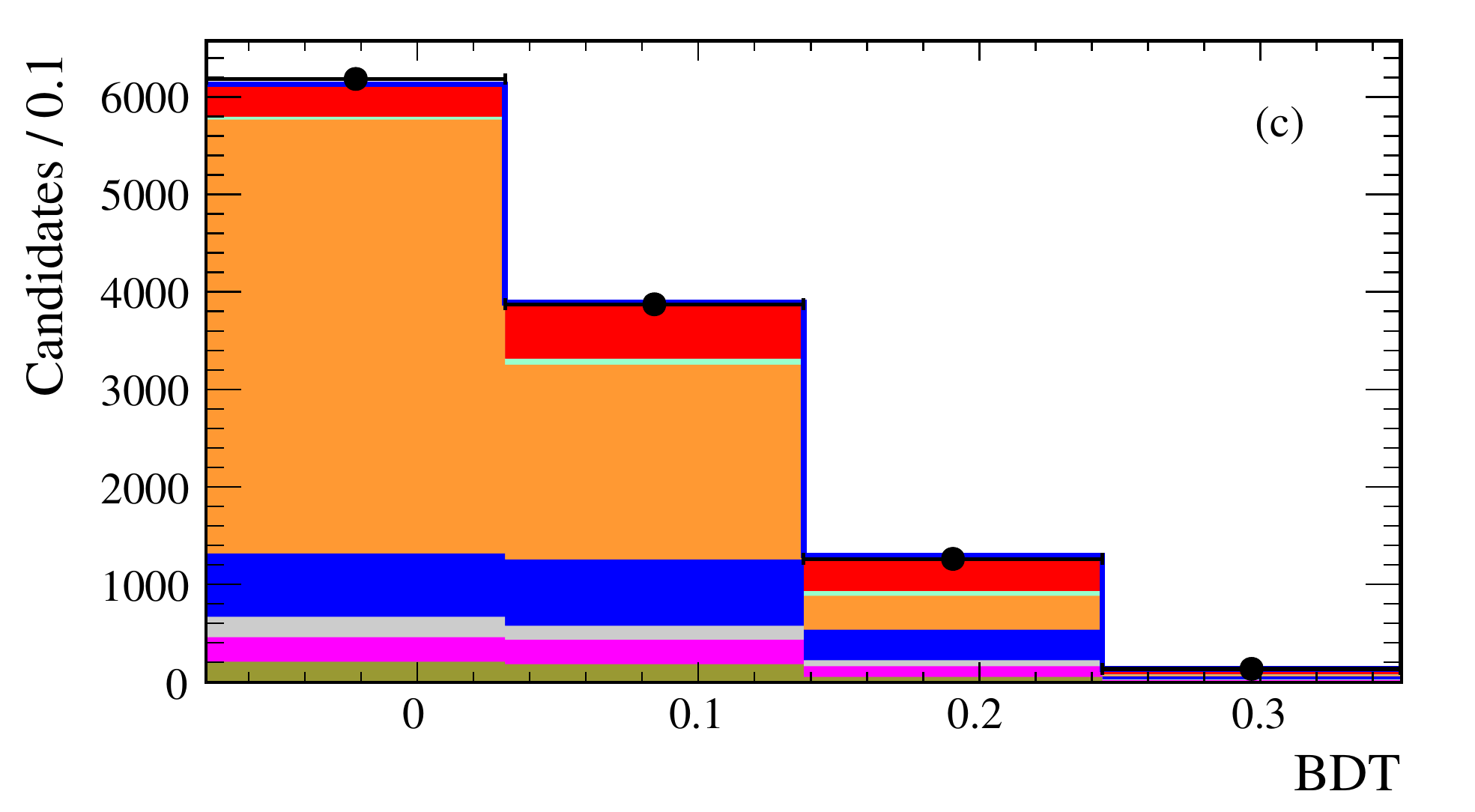}
\vspace{-0.2cm}
\caption{Fit projections of the three-dimensional fit of the 3$\pi$ decay time, \qsq, and BDT output distributions from \lhcb's hadronic \RDst measurement~\cite{Aaij:2017deq}.}
\label{fig:RDsthad}
\end{figure*}

In the hadronic measurement of \RDst~\cite{Aaij:2017uff,Aaij:2017deq} the \taum lepton is reconstructed with three charged pions in the final state. 
Instead of the $\Bzb \to \Dstarp \mun \neumb$
decay mode, this analysis uses the decay $\Bzb\to\Dstarp\pim\pip\pim$ as a normalisation channel. It then measures the ratio 
$\mathcal{K}(\Dstar)$, which is defined as:
\begin{equation}
\mathcal{K}(\Dstar) = \frac{\BR(\Bzb\to\Dstarp\taum\neutb)}{\BR(\Bzb\to\Dstarp\pim\pip\pim)} \, .
\end{equation}

To convert this value to \RDst, $\mathcal{K}(\Dstar)$ is multiplied by the ratio of the branching ratios of the $\Bzb\to\Dstarp\pim\pip\pim$ 
and $\Bzb\to\Dstarp\mun\neumb$ decays, which are taken as external inputs from HFLAV average:
\begin{equation}
\RDst = \mathcal{K}(\Dstar) \times \left( \frac{\BR(\Bzb\to\Dstarp\pim\pip\pim)}{\BR(\Bzb\to\Dstarp\mun\neumb)} \right) \, .
\end{equation}

This analysis benefits from the well-defined \taum decay vertex which is downstream from the \B decay vertex, and 
suppresses backgrounds by exploiting this topology. For the signal channel a template fit is performed in three variable: the decay time of the 
three pions ($t_{\tau}$), \qsq, and the output of a boosted decision tree (BDT). This BDT is used to suppress backgrounds coming from doubly-charmed
$\Bz\to\Dstarm\Dsp X$ decays, where $X=\Bp,\Bz,\Bs$. Projections of the fits for each of these variables are shown in Fig.~\ref{fig:RDsthad}.
The analysis yields a value of 
\begin{eqnarray*}
\mathcal{K}(\Dstar) = 1.93 \pm 0.12\stat \pm 0.17\syst \, .
\end{eqnarray*}

Recently HFLAV updated the external input of the average of the
measurements of $\BR(\Bz\to\Dstarm\ellp\neul)$, which changed from $(4.88\pm0.01\stat\pm0.10\syst)\%$ to 
$(5.08\pm0.02\stat\pm0.12\syst)\%$. The change is largely due to the decision to no longer average over the \Dstarp and 
\Dstarz decays, resulting in the exclusion of measurements combining these states. Moreover, the new average includes the latest
\belle measurement~\cite{Abdesselam:2018nnh}.

Using the updated HFLAV average, \lhcb's measurement of \RDst using the hadronic \taum decay yields a value of:
\begin{eqnarray*}
\RDst = 0.280 \pm 0.018\stat \pm 0.029\syst \, .
\end{eqnarray*}
This is in agreement with the SM within $1\sigma$.

\subsection{Muonic \RJpsi}

\begin{figure*}
  \centering
  \setbox1=\hbox{\includegraphics[width=0.33\textwidth]{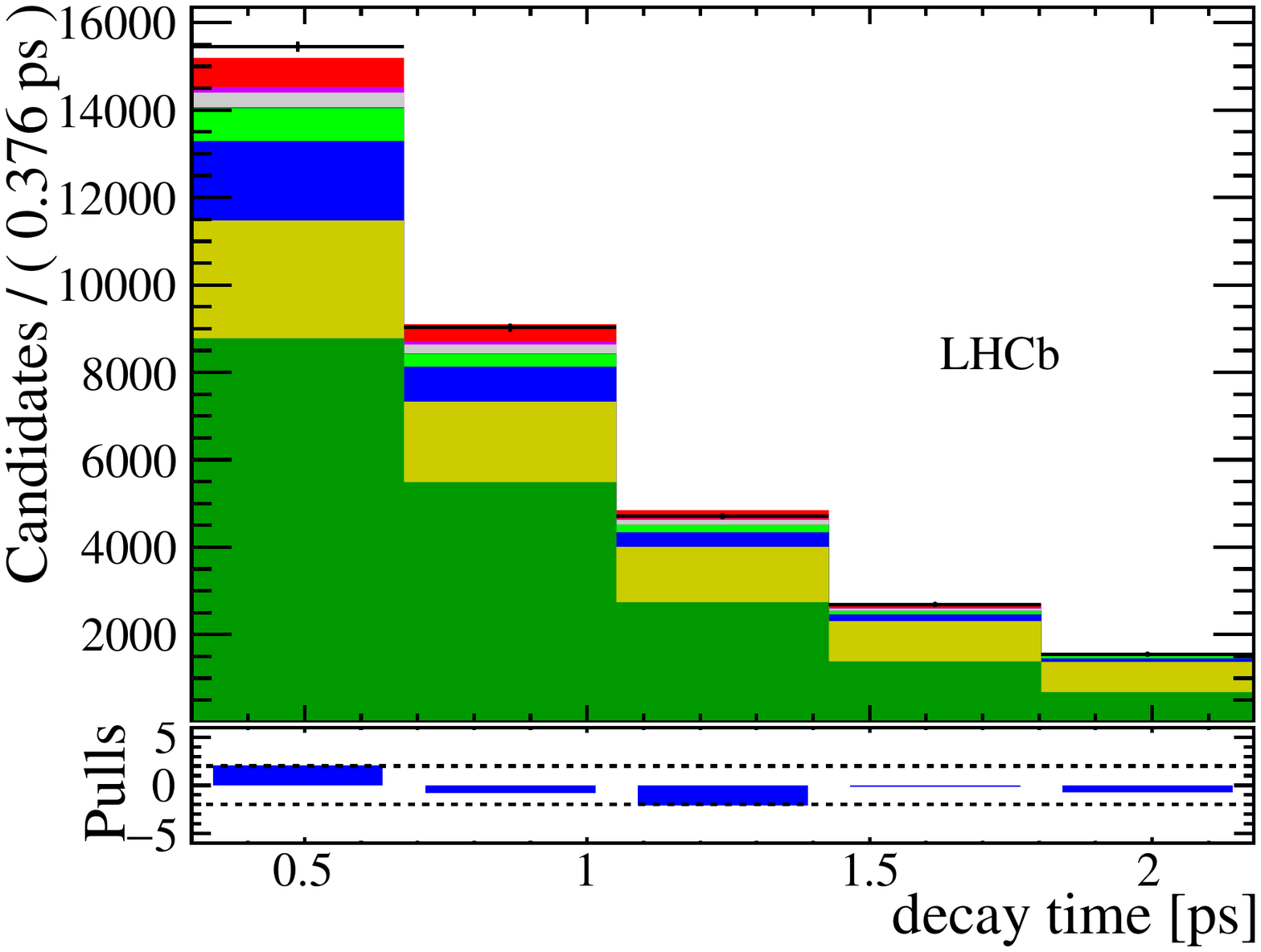}}  
  \includegraphics[width=0.32\textwidth]{figs/figure3b.pdf}\llap{\makebox[3.55cm][l]{\raisebox{2.8cm}{\includegraphics[height=0.85cm]{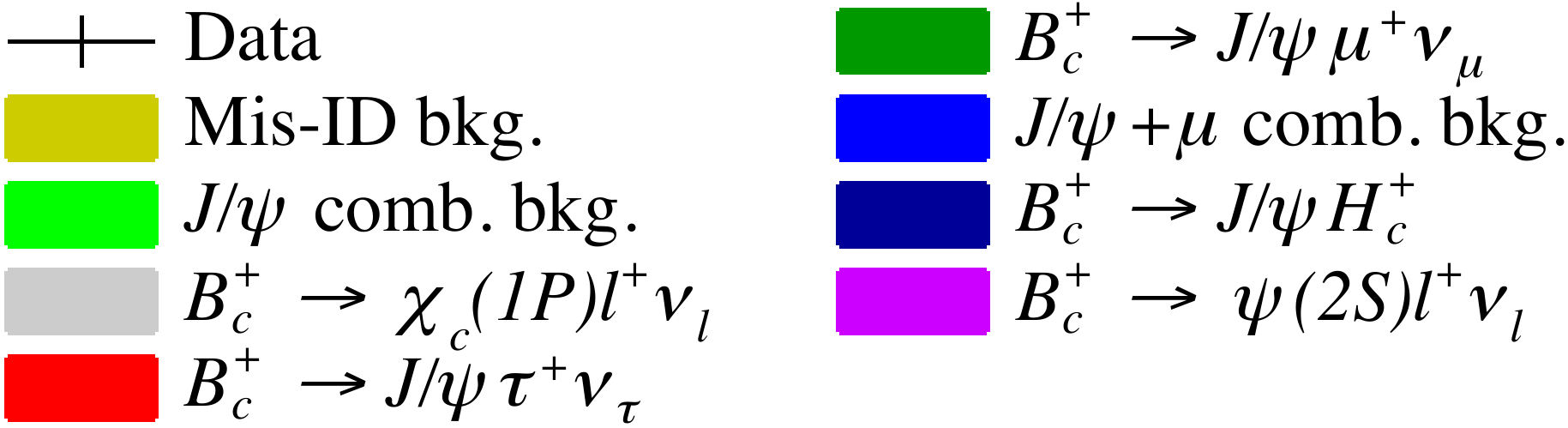}}}}
  \includegraphics[width=0.32\textwidth]{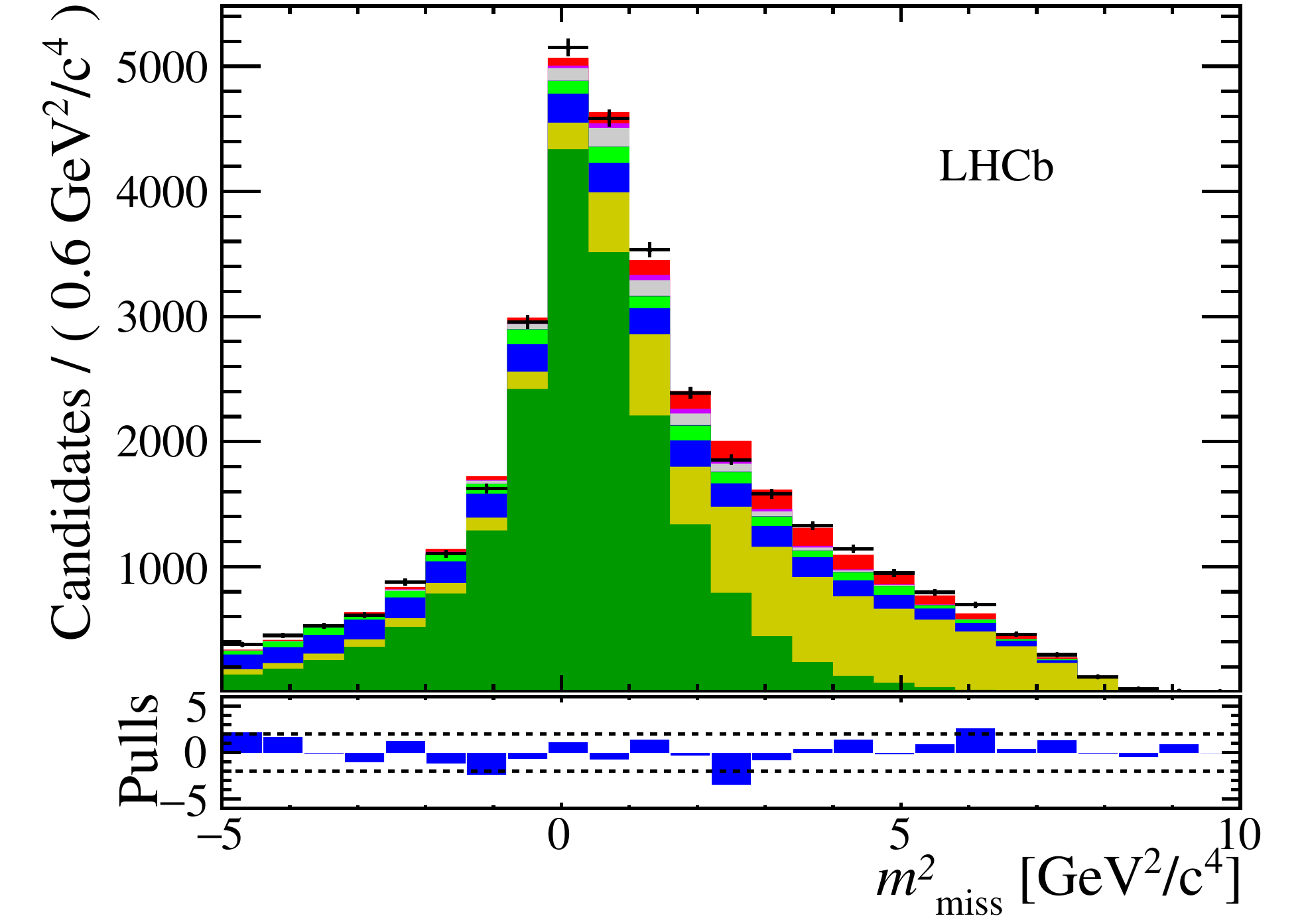}
  \includegraphics[width=0.32\textwidth]{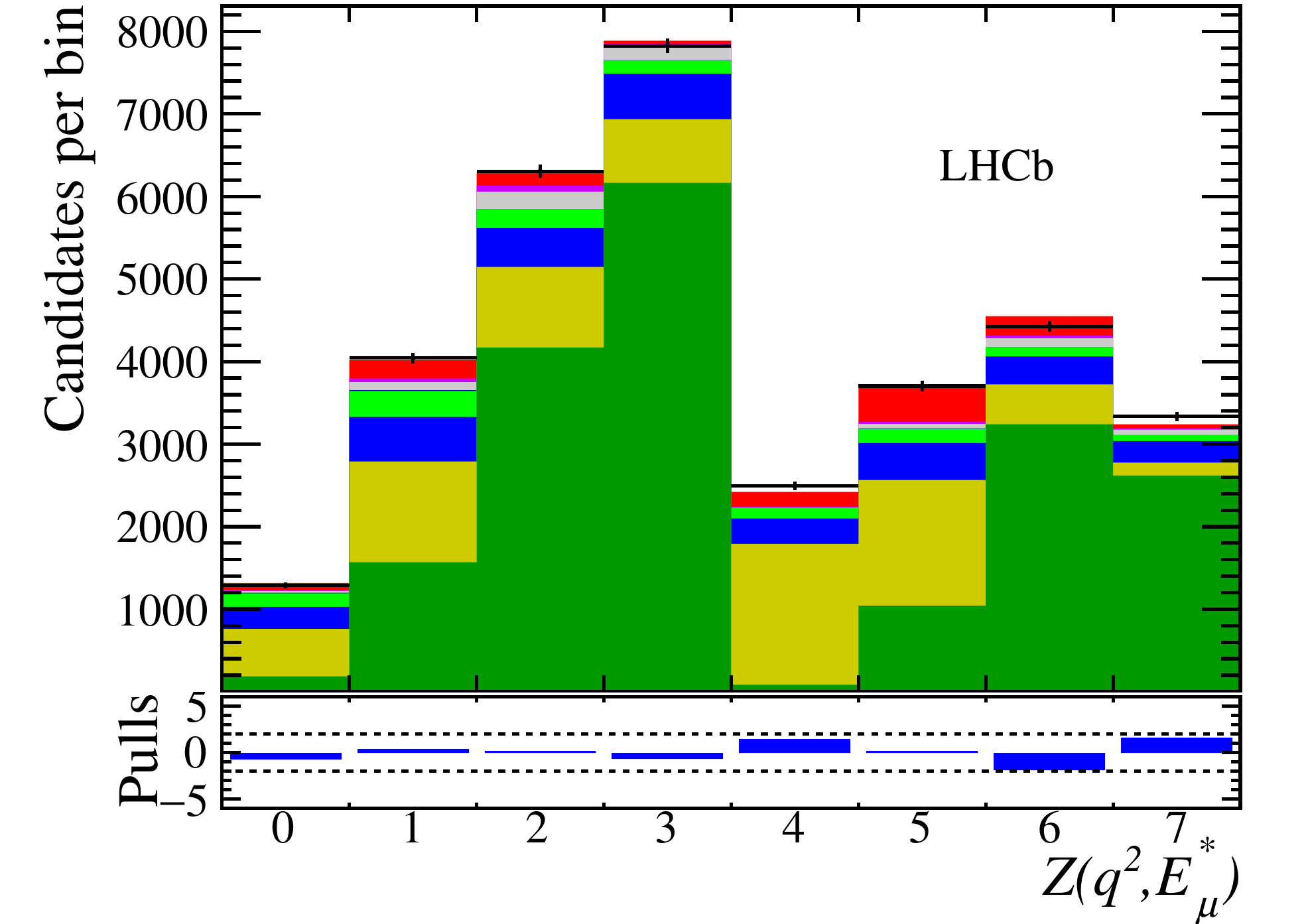}
  \caption{\label{fig:RJpsit_fits} Fit projections of the $\tau$ decay time (left), \mmiss (middle) and $Z(\qsq,E_{\mu}^*)$ variables in the measurement of \RJpsi from \lhcb~\cite{Aaij:2017tyk}.}
\end{figure*}

The latest measurement from \lhcb presented here~\cite{Aaij:2017tyk} studies the ratio in a different decay mode, namely
\begin{equation}
\RJpsi = \frac {\Bc\to\jpsi\taup\neut}{\Bc\to\jpsi\mup\neum} \, .
\end{equation}
The \taum lepton is reconstructed in the muonic decay mode and also in this analysis the signal and normalisation channel are 
distinguished in a three-dimensional templated fit based on the \taum decay time, \mmiss, and the variable $Z(\qsq,E_{\mu}^*)$, 
which is a combination of the \qsq and $E_{\mu}^*$ variables. The same boost approximation is used as in the muonic \RDst 
analysis. Projections of the fit output are shown in Fig.~\ref{fig:RJpsit_fits}.

The analysis yields a value of 
\begin{eqnarray*}
\RJpsi = 0.71 \pm 0.17\stat \pm 0.18\syst \, ,
\end{eqnarray*}
where one of the largest systematic uncertainties comes from the limited knowledge on the form factors of the 
$\Bc\to\jpsi\ellp\neul$ decays.
These are currently fit from data but can be significantly improved with new lattice calculations.
\RJpsi is compatible with the SM within 2$\sigma$.

\section{Latest measurement from Belle}
The latest measurement of LFU in charged-current \B decays of the Belle collaboration~\cite{Abdesselam:2019dgh} simultaneously measures \RD and \RDst. 
It analyses the full \FourS sample recorded by the Belle detector,
consisting of $772\times10^6$ \B\Bb events. It uses a semileptonic tag, meaning that the 
other \B meson in the event is reconstructed in the semileptonic decay $\B\to\D^{(*)}\ell\neul$, where $\ell=e,\mu$. 
Since the previous \RDst analysis with a semileptonic tag~\cite{Sato:2016svk}, the tagging algorithm has been 
extended with more reconstruction channels and now uses a BDT resulting in a sample with higher signal purity.

In order to make sure the tag \B meson does not decay with a \taum lepton in the final state, a cut on the variable $\cos \theta_{\B,\D^{(*)}\ell}$ is applied, 
where $\cos \theta_{\B,\D^{(*)}\ell}$ is the cosine of the angle between the momentum of the \B meson 
and the $\D^{(*)}\ell$ combination in the \FourS rest frame. This variable is reconstructed assuming that there is only one massless unreconstructed particle 
(neutrino) in the decay and it is defined as:
\begin{equation}
\cos \theta_{\B,\D^{(*)}\ell} \equiv \frac{2 E_{\rm beam}E_{\D^{(*)}\ell}-m_B^2-m_{\D^{(*)}\ell}^2}{2|p_B||p_{\D^{(*)}\ell}|} \, ,
\end{equation}
where $E_{\rm beam}$ is the energy of the beam, and $E_{\D^{(*)}\ell}$, $m_{\D^{(*)}\ell}$, and $p_{\D^{(*)}\ell}$ are the energy, 
mass and momentum of the $\D^{(*)}\ell$
system, respectively. The variable $m_B$ is the nominal \B meson mass, and $p_B$ the \B meson momentum.

The \B mesons are reconstructed in the $\Dp\ellm$, $\Dz\ellm$, $\Dstarp\ellm$ and $\Dstarz\ellm$ decays, which increases the signal yields compared to the previous
semileptoni-tag analysis by Belle~\cite{Sato:2016svk} because now both \Bp and \Bz decays are studied, rather than only \Bz decays. 
The \Dstar mesons are reconstructed as $\Dstarp\to\Dz\pip$, $\Dstarp\to\Dp\piz$, and $\Dstarz\to\Dz\piz$.
The \Dz and \Dp mesons are reconstructed in various final states with kaons and pions, adding up 30\% and 22\% of the total \Dz and \Dp branching fractions, respectively.
To reduce backgrounds, the \D candidates are required to be in a mass window within 15\mevcc of their nominal mass, although this mass window is extended for 
\D mesons with a \piz in the final state due to the worse resolution for these events. In every event, the two \B mesons are required to have opposite flavour to reduce combinatorial backgrounds.

\begin{figure}[htb]
  \includegraphics[width=0.9\columnwidth]{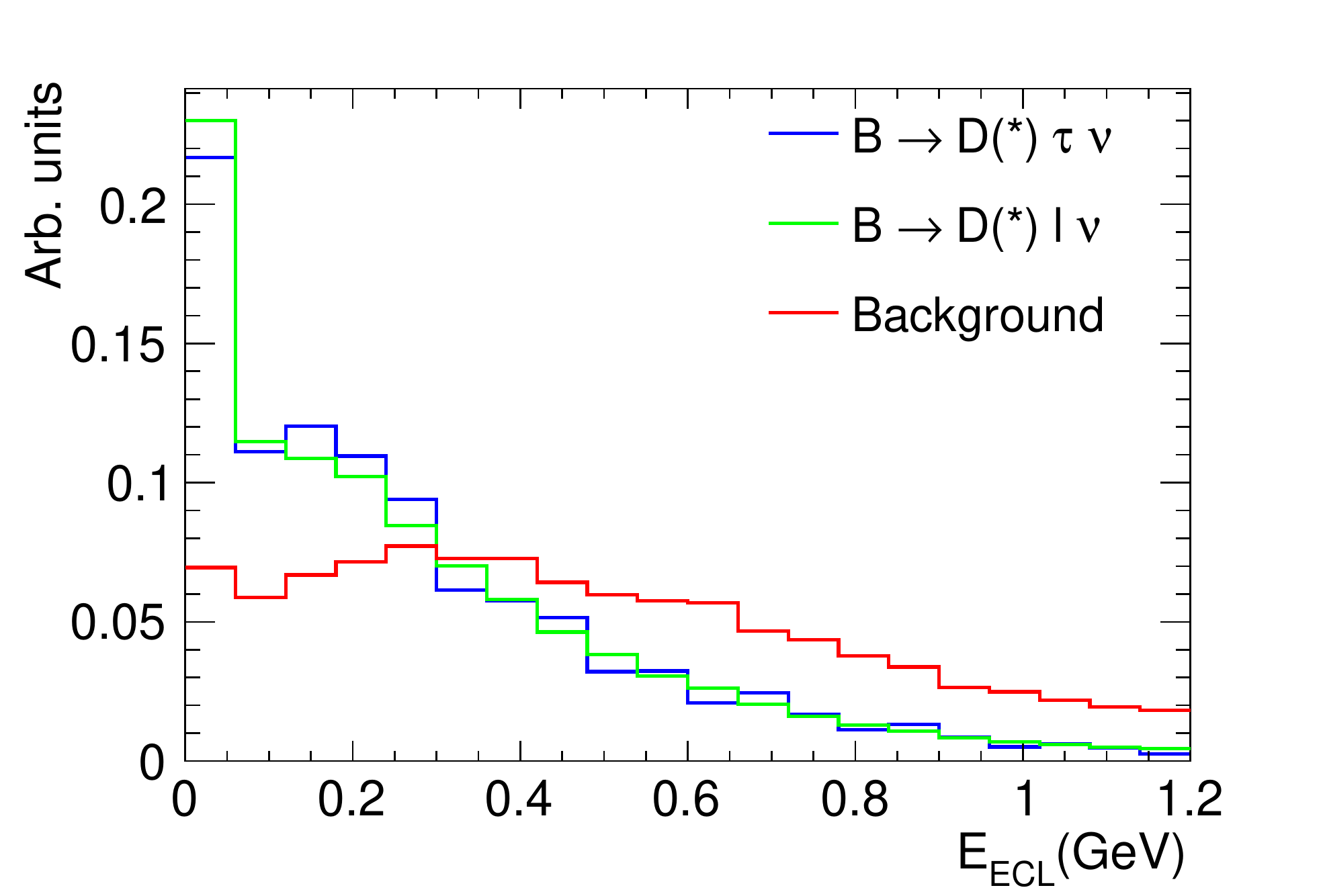}
    \vspace{-0.4cm}
  \caption{Distributions of the $E_{\rm ECL}$ variable used in \belle's semileptonic tag measurement~\cite{Abdesselam:2019dgh}. The distributions of the signal, normalisation and 
  background samples are taken from simulation and normalised to unity.} 
  \label{fig:th_MC_eecl}
\end{figure}

For each of the four samples, 
a two-dimensional template fit is performed to distinguish signal, normalisation and background yields. 
The two parameters used to fit in are $E_{\rm ECL}$ and \verb+class+. The former is 
the energy deposited in the calorimeter which is not associated with reconstructed particles. This energy, which is restricted to be less than 1.2\gev, peaks at zero for the signal and normalisation channels, while it has a
reasonably flat distribution for the background components, as illustrated in Fig.~\ref{fig:th_MC_eecl}. The \verb+class+ variable is the output of a BDT based 
on the visible energy $E_{\rm vis}$, \mmiss, and $\cos\theta_{\B,\D^{(*)}\ell}$. No further selection is applied to this variable.

The fits are performed simultaneously on the four samples and consists of templates for the following components: 
\begin{itemize}
\item $\D^{(*)}\tau\neut$  ,
\item $\D^{(*)}\ell\neul$  ,
\item $\D^{**}\ell\neul$, where $D^{**} = \D_1,\D_2^*,\D_1^{'}, \D_0^*$  ,
\item feeddown from  $\D^{*}\ell\neul$ to $\D\ell\neul$ decays ,
\item fake $\D^{(*)}$, fixed in the fit
\item other backgrounds, fixed in fit
\end{itemize}
The fit PDFs are based on simulation samples which have a luminosity of ten times the total collected \B\Bb luminosity for the signal and normalisation channels, and five times for
the $\D^{**}$ states. 
To get an estimate of the feed down, the result of the $\Dstar\ell$ ($\Dstar\tau$) fit is used to constrain this component in the $\D\ell$ ($\D\tau$) fit.
The number of fake $\D^{(*)}$ decays is determined from the $\Delta m = m_{\Dstar}-m_{\D}$ sidebands and the yields of the other backgrounds are fixed to their 
simulation expectation value.

\begin{figure*}[tbp]
\includegraphics[width=0.87\columnwidth]{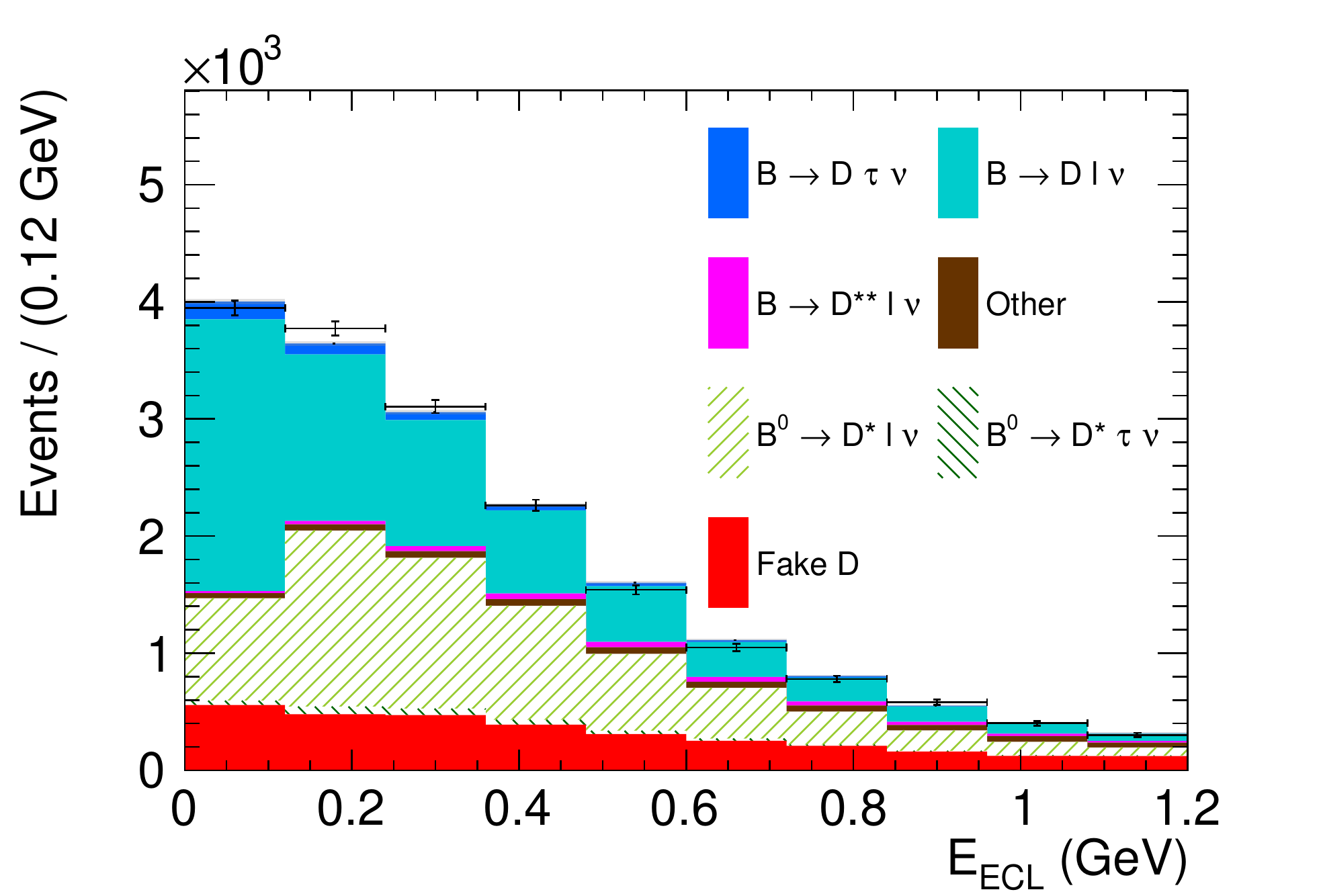}
\includegraphics[width=0.87\columnwidth]{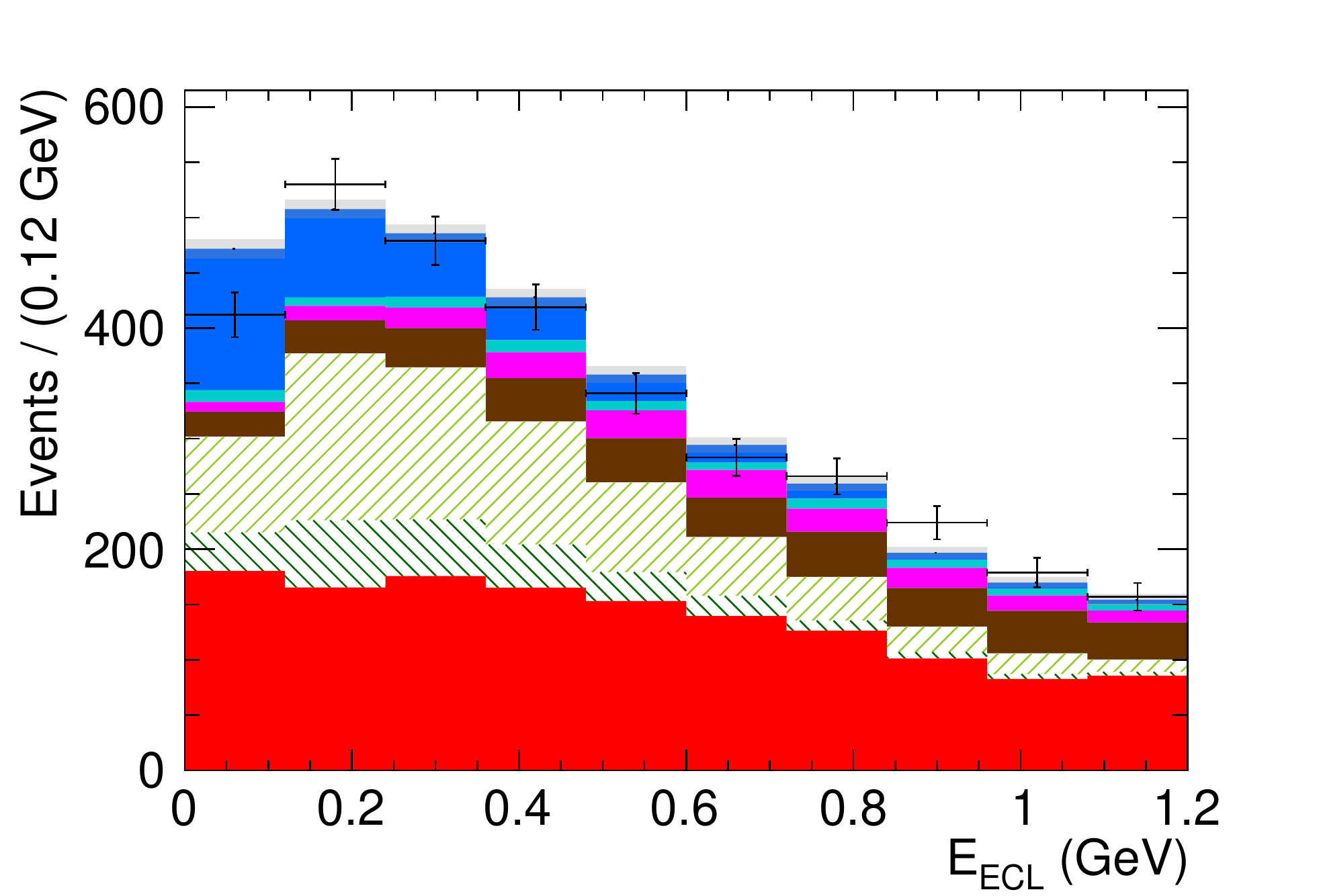}
\includegraphics[width=0.87\columnwidth]{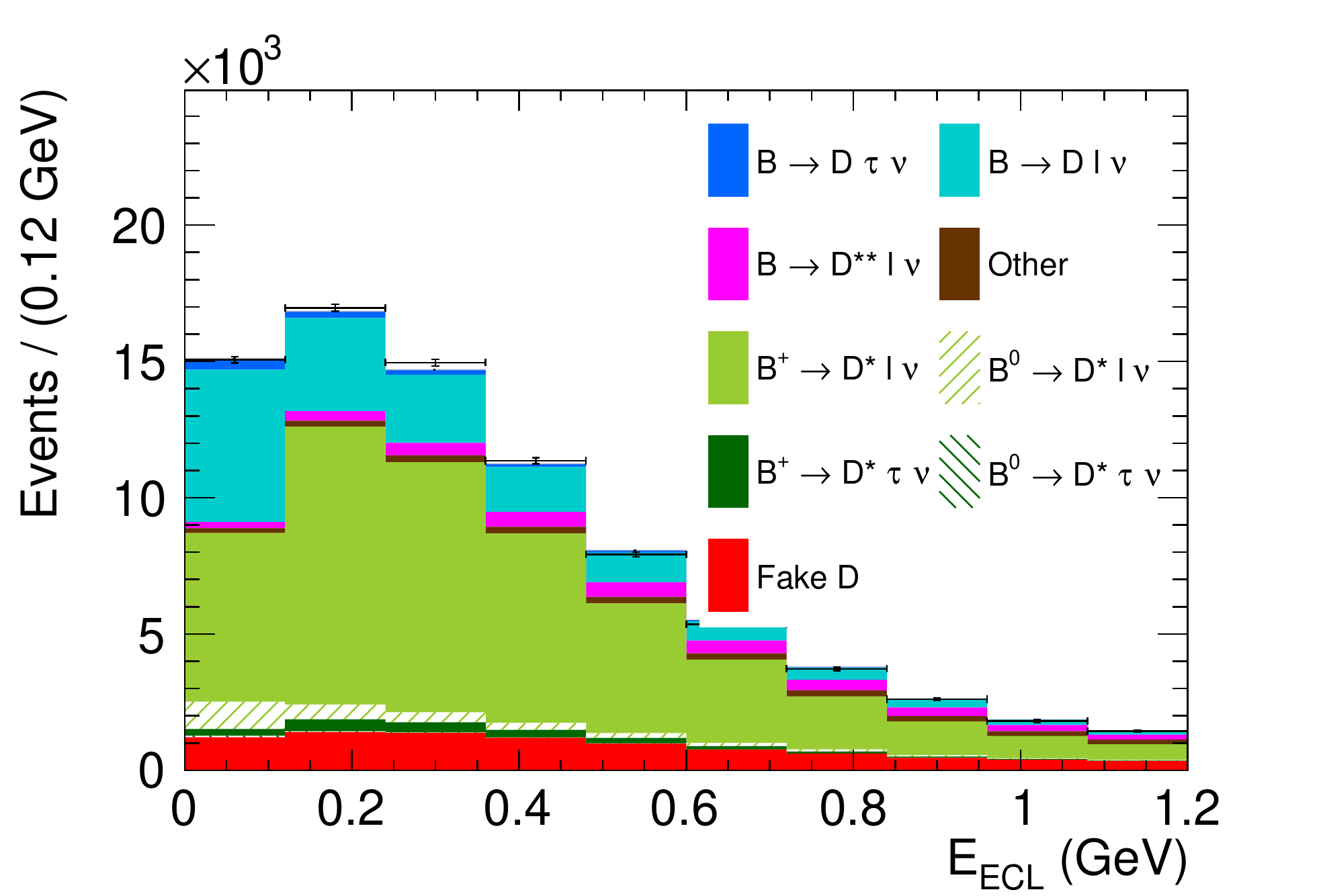}
\includegraphics[width=0.87\columnwidth]{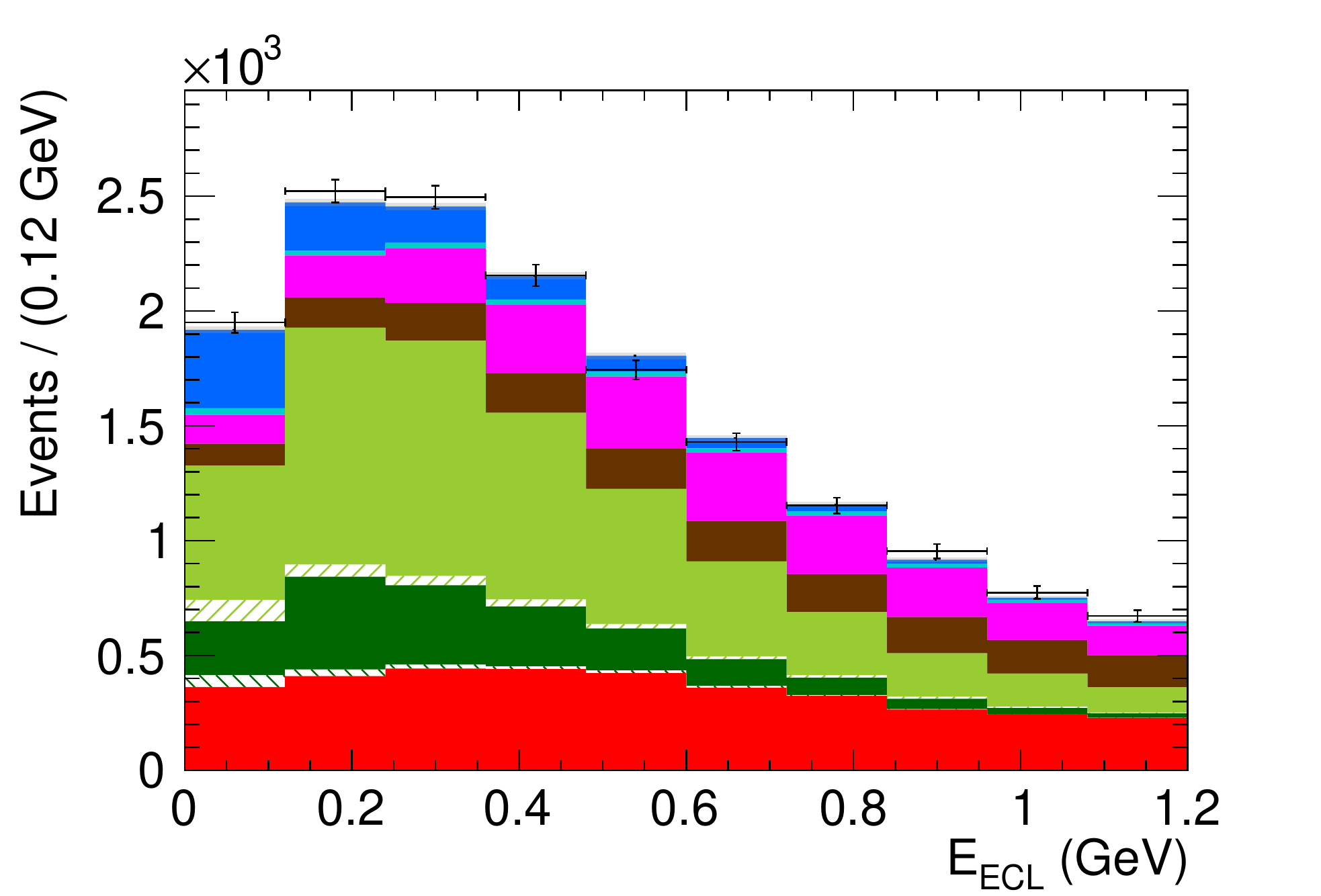}
\caption{
Fit projections of the $D^+\ell^-$ (top) and $D^0\ell^-$ (bottom) samples as a function of $E_{\rm ECL}$~\cite{Abdesselam:2019dgh}. The plots on the left show the full classifier region, while the plots on the right are the signal region, defined by the selection \texttt{class} $>0.9$.}
\label{fig:results_Dmodes}
\end{figure*}

\begin{figure*}[tbp]
\includegraphics[width=0.87\columnwidth]{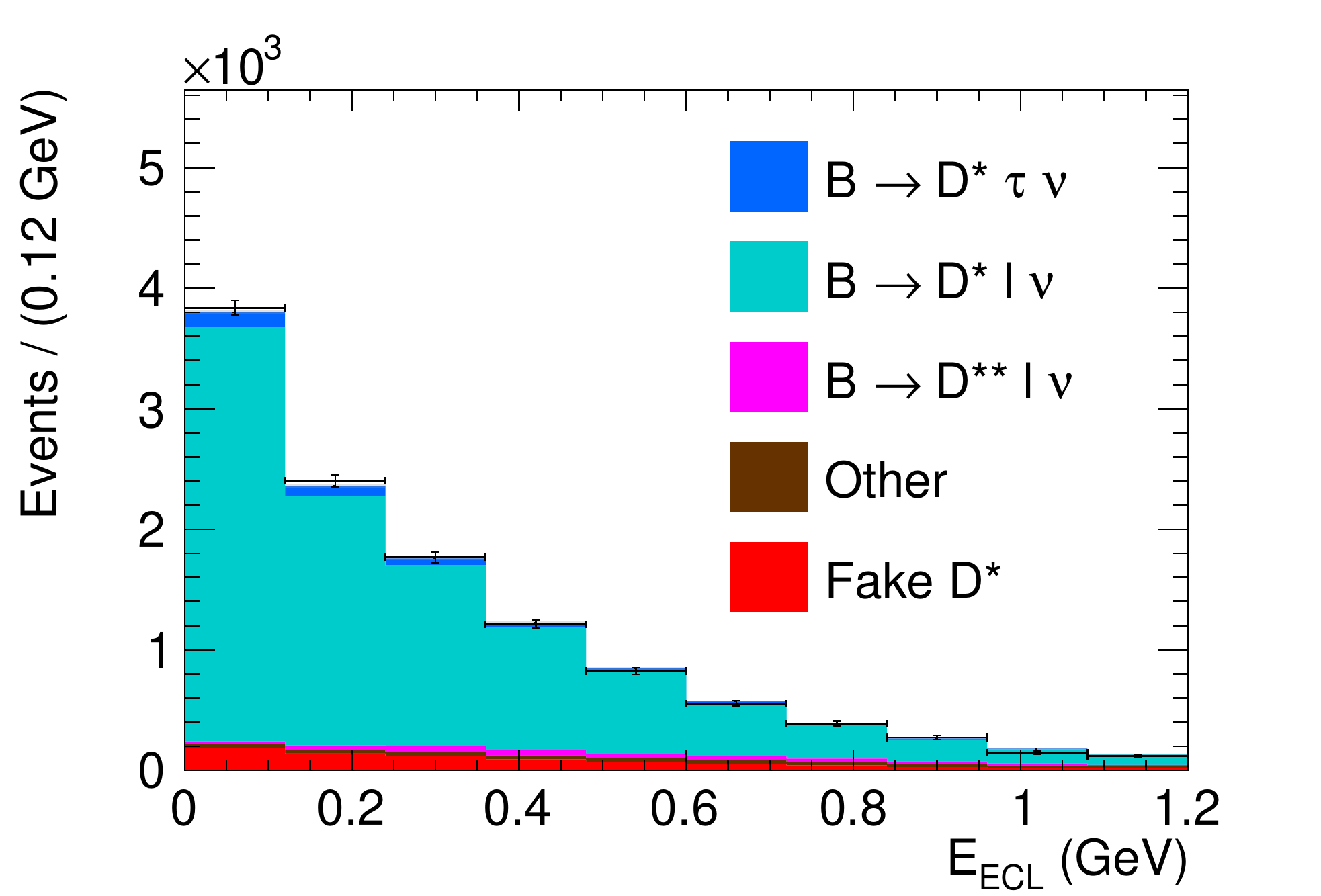}
\includegraphics[width=0.87\columnwidth]{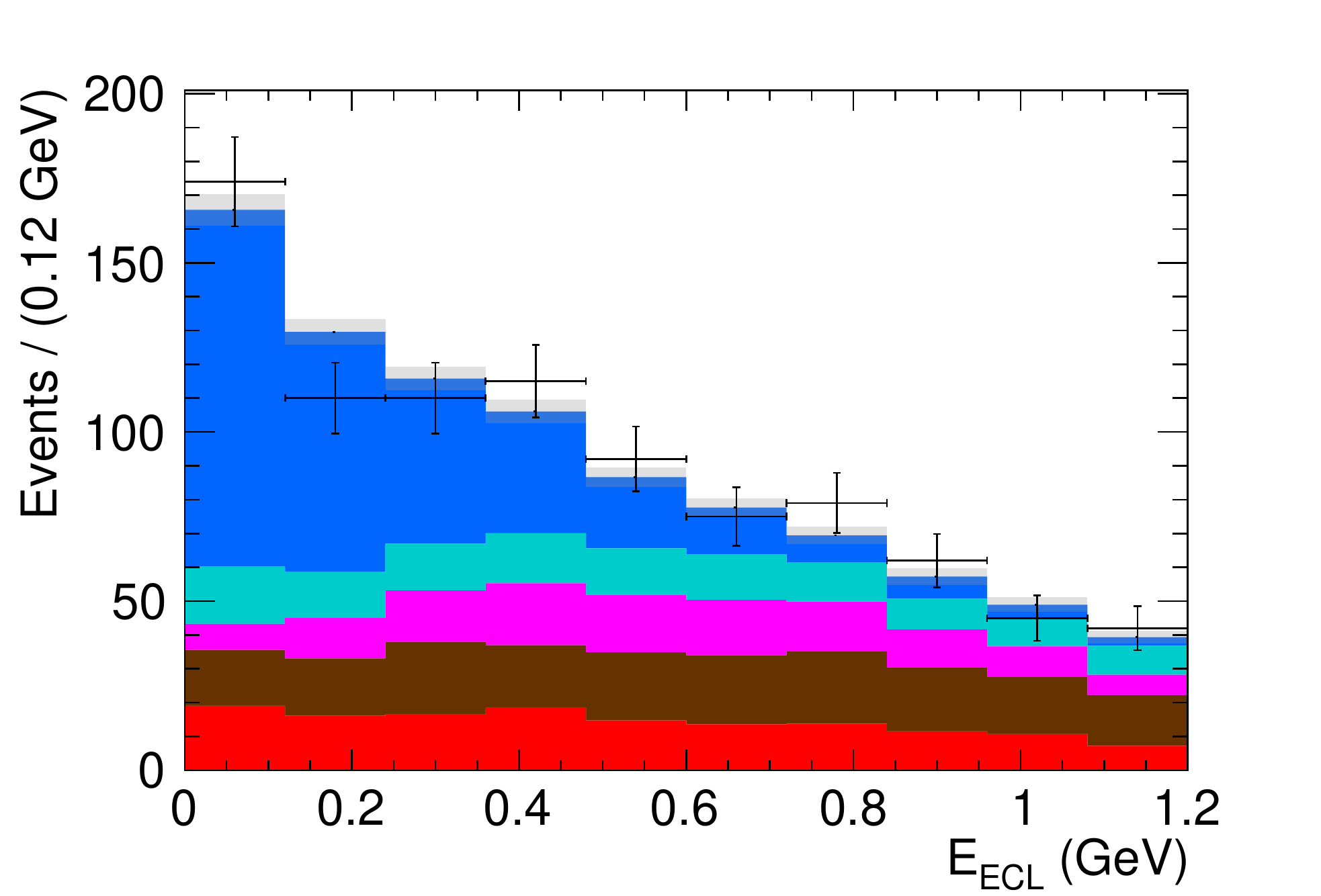}
\includegraphics[width=0.87\columnwidth]{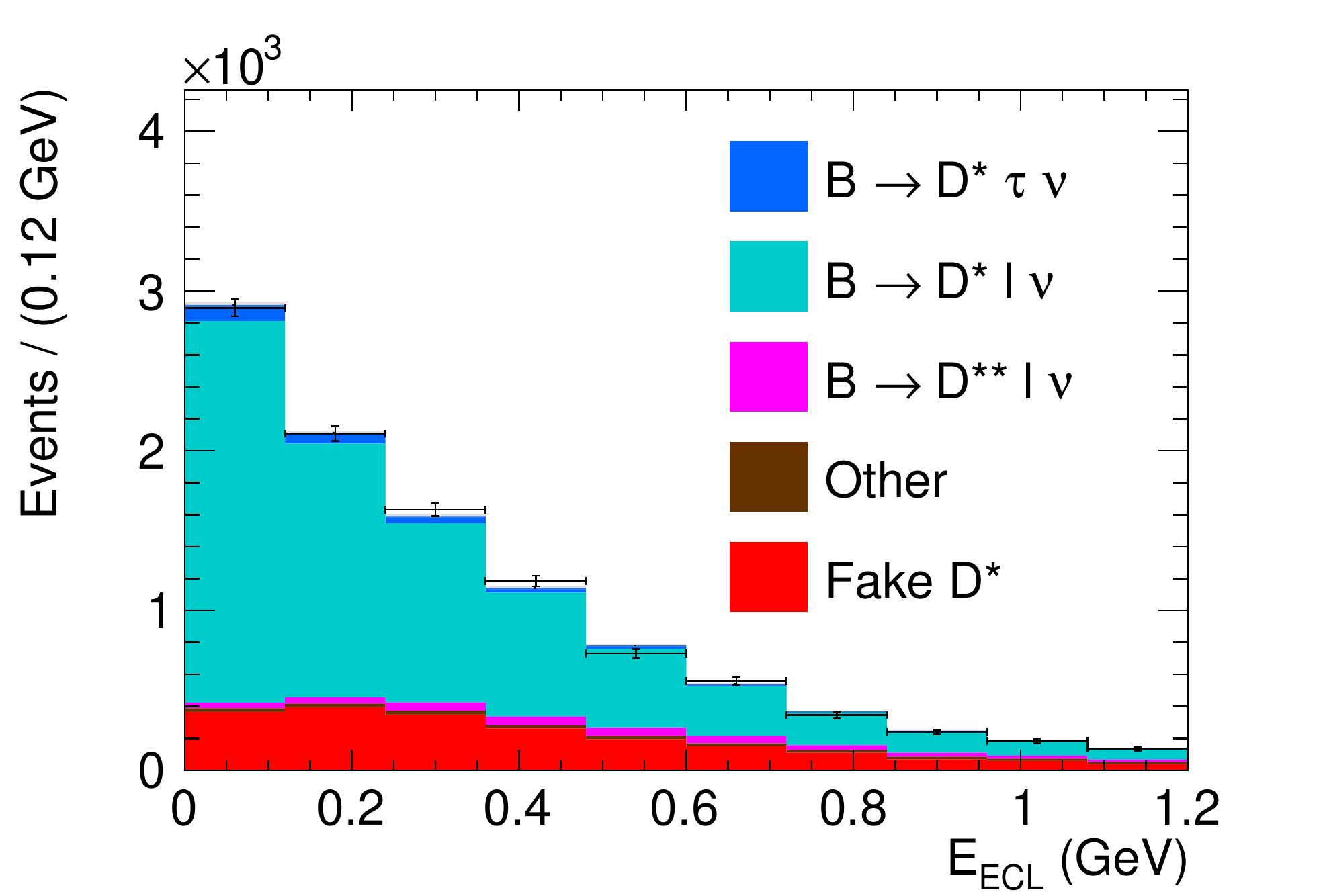}
\includegraphics[width=0.87\columnwidth]{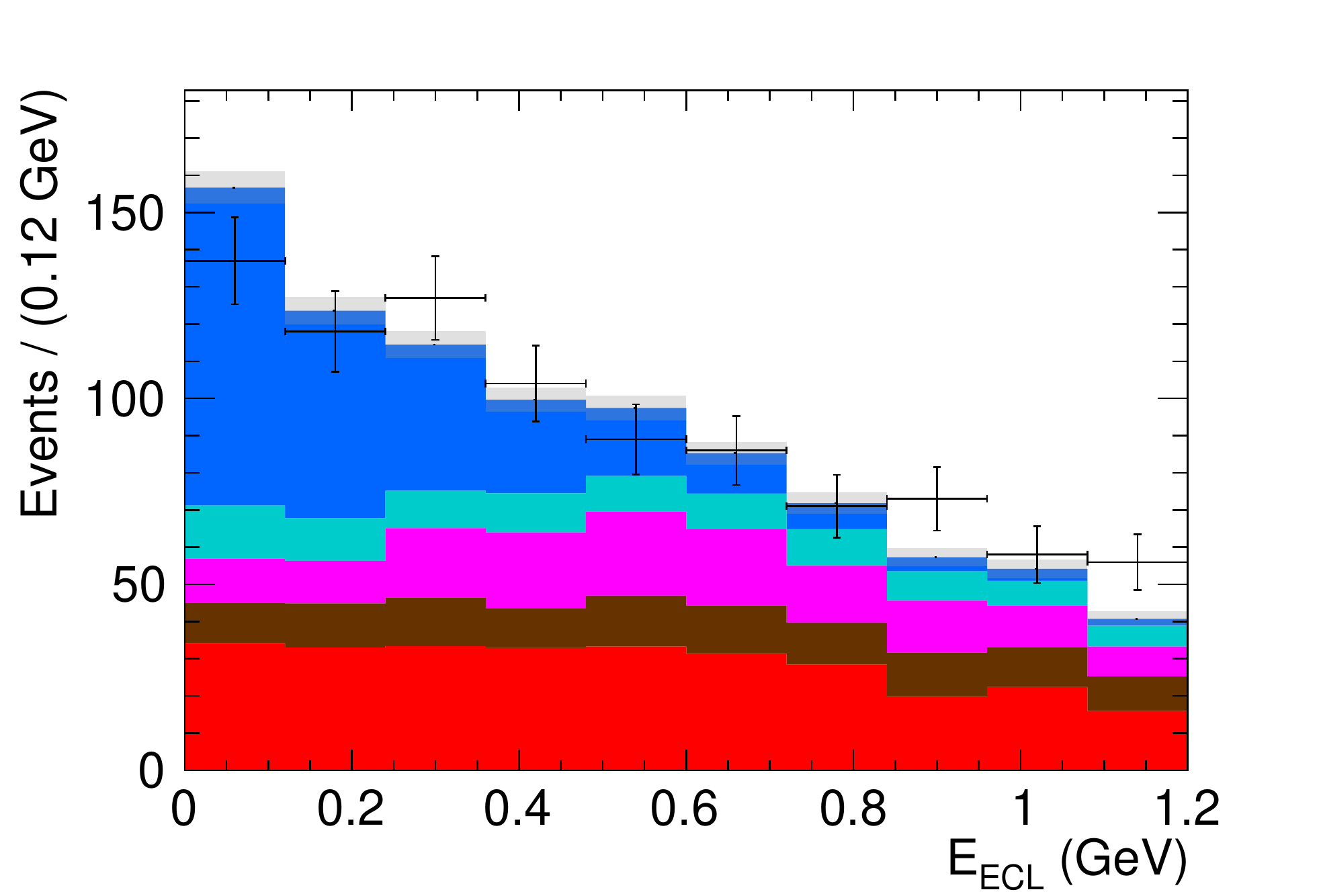}
\caption{Fit projections of the $D^{\ast +}\ell^-$ (top) and $D^{\ast 0}\ell^-$ (bottom) samples as a function of $E_{\rm ECL}$~\cite{Abdesselam:2019dgh}. The plots on the left show the full classifier region, while
the plots on the right are the signal region, defined by the selection \texttt{class} $>0.9$.}
\label{fig:results_Dstmodes}
\end{figure*}

Fit projections of the $D^+\ell^-$ and $D^0\ell^-$ samples are shown in Fig.~\ref{fig:results_Dmodes}. The blue signal samples are hardly visible in the plots
on the left showing the full classifier region. To illustrate the region associated with signal, also the fit results for the region with \texttt{class} $>0.9$ are shown 
in Fig.~\ref{fig:results_Dmodes} (right). Here, the signal is much more visible, and the contribution of the normalisation channel is reduced. 
Fig.~\ref{fig:results_Dstmodes} shows similar plots, but for the $\Dstarp\ellm$ and $\Dstarz\ellm$ samples. 

Finally, $\mathcal{R}(D^{(*)})$ can be calculated using the following expression:
\begin{equation}
\mathcal{R}(D^{(*)}) = \frac{1}{2 \mathcal{B}(\taum \to \ellm \neulb \neut)} \cdot \frac{\varepsilon_{\rm norm}}{\varepsilon_{\rm sig}} \cdot \frac{N_{\rm sig}}{N_{\rm norm}} \, ,
\end{equation}
where $\varepsilon_{\rm sig(norm)}$ and $N_{\rm sig(norm)}$ are the detection efficiency and fitted yields of the signal and normalisation modes, respectively.
$\mathcal{B}(\taum \to \ellm \neulb \neut)$ is the world average for $\ell=e,\mu$. The efficiencies are taken from simulation samples, which are
corrected to resemble the data more closely by applying correction factors. One of the largest corrections is to the 
lepton identification efficiency, which is corrected separately for electrons and muons. The efficiencies are corrected based on their kinematical 
dependence using control samples of
$\epem\to\epem\ellell$ and $\jpsi\to\ellell$ decays.

The analysis measures values of
\begin{eqnarray*}
\RD    &= 0.307 \pm 0.037\stat \pm 0.016\syst \, , \\
\RDst &= 0.283 \pm 0.018\stat \pm 0.014\syst \, ,
\end{eqnarray*}
where the correlation between the statistical uncertainties and between the systematic uncertainties is $-0.53$ and $-0.52$, respectively.
These are the most precise measurements of \RD and \RDst to date and they are in agreement with the SM within 0.2$\sigma$ and 1.1$\sigma$, respectively.
The combined result agrees with the SM prediction within 1.2$\sigma$.
The largest contributions to the systematic uncertainties come from the limited size of the simulation sample, and the knowledge on the reconstruction efficiency. 

\section{Conclusions}
\begin{figure}[tb]
\includegraphics[width=\linewidth]{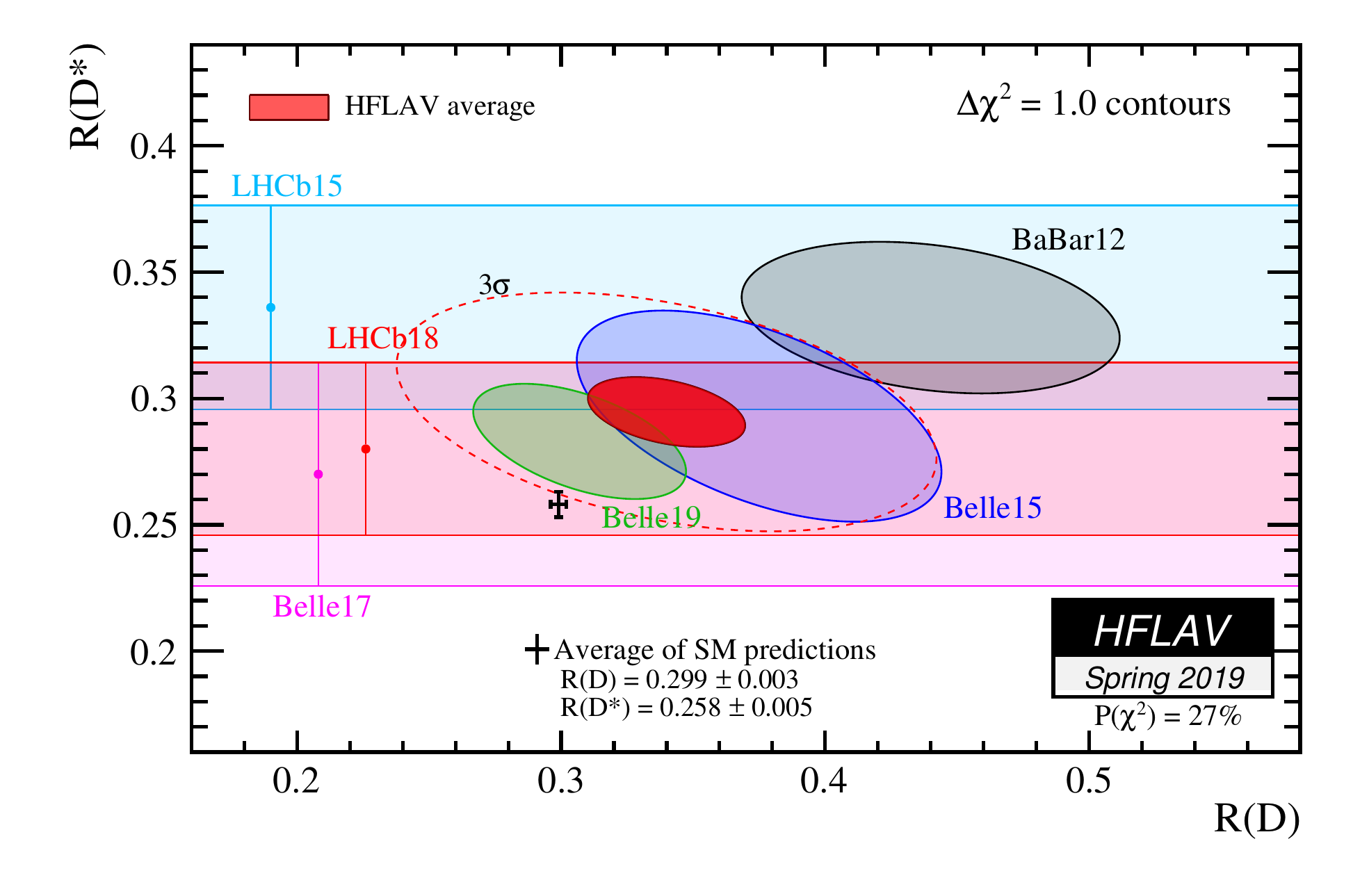} 
\vspace{-0.4cm}
\caption{HFLAV average of all measurements of \RD and \RDst, updated with the results of the \belle collaboration presented in these proceedings. The red ellipse 
shows the combined average and the data point is the SM prediction, showing a discrepancy of 3.1$\sigma$. This plot is retrieved from the HFLAV website~\cite{HFLAV16}, using inputs from~\cite{Lees:2012xj,Lees:2013uzd,Huschle:2015rga,Aaij:2015yra,Hirose:2016wfn,Hirose:2017dxl,Aaij:2017uff,Aaij:2017deq,Abdesselam:2019dgh}.}
\label{fig:HFLAV}
\end{figure}

The HFLAV group produced new averages of all measurements of \RD and \RDst, including the latest result from Belle and the
update of the external input for \lhcb's hadronic \RDst measurement. The current averages are:
\begin{eqnarray*}
\RD    &= 0.349 \pm 0.027\stat \pm 0.015\syst \, , \\
\RDst &= 0.298 \pm 0.011\stat \pm 0.007\syst \, .
\end{eqnarray*}
The combination of all measurements of \RD and \RDst, which is shown in Fig.~\ref{fig:HFLAV}, yields a 3.1$\sigma$ discrepancy 
with the SM.

Many new measurements of LFU in charged-current \B decays in \lhcb are on their way. Work is ongoing on updates of the measurements presented in these proceedings, 
including the extension of the muonic \RDst measurement to the combination of \RD-\RDst. Additionally, other decay channels are being studied,
these measure the ratios $\mathcal{R}(\Dp)$, $\mathcal{R}(\Lc)$, $\mathcal{R}(\Ds)$, $\mathcal{R}(\proton\proton)$. They are analysed both in muonic and hadronic 
decay mode of the \taum lepton, and, depending on the measurement, use the Run 2 as well as the Run 1 dataset. These measurements will shed new light on the current
discrepancy with the SM.
Finally, the large datasets that will be collected by the \lhcb upgrade~\cite{Bediaga:2018lhg} and Belle II~\cite{Kou:2018nap} experiments will allow measurements 
of LFU in charged-current \B decays to be precise enough to confirm LFU breaking if the central values remain the same as the current best-fit values. 

\bigskip 
\bibliography{main}

\end{document}